\documentclass[showpacs,preprint]{revtex4}
\usepackage{graphicx}
\usepackage{subfigure}
\usepackage{amssymb}
\usepackage{amsmath}
\usepackage{bm}
\usepackage{latexsym}
\usepackage{hyperref}

\newcommand{\Tr}{{\text Tr}}
\begin{document}
\title{Level crossing and quantum phase transition of the XY model}
\author{Jun Jing\footnote{Email: jingjun@sjtu.edu.cn}, H. R. Ma }
\affiliation{Institute of Theoretical Physics,
Shanghai Jiao Tong University\\
800 DongChuan Road, MinHang, Shanghai 200240, China}
\date{\today}

\begin{abstract}
The ground state of a one-dimensional spin-1/2 chain with periodical
boundary condition in the Heisenberg XY model is investigated. We
consider the spatial correlation and concurrence between any
nearest-neighbor pair of spins under the conditions of different
coupling strength, anisotropic parameter and magnitude of a
transverse field. Quantum phase transitions due to the competition
between coupling and alignment which cause the abrupt changes of the
correlation and concurrence are observed. The transition are direct
results of the level crossing.

\end{abstract}
 \maketitle

\section{Introduction}

The quantum Heisenberg XY model, which was first intensively studied
by Lieb et al in the early 1960s \cite{Lieb}, has aroused new
interests in recent years. It was taken to be a potential candidate
model in the practical applications in quantum computation
\cite{Zheng, You, Raussendorf} and quantum information
\cite{Imamoglu,Burkard}. The Heisenberg model is not only a
theoretical model with many interesting physics in it, but also
realized in recent years in laboratory in physical systems such as
quantum dots \cite{Loss}, nuclear spins \cite{Kane} and electronic
spins \cite{Vrijen}. Many theoretical researchs \cite{Connor, Wang}
had been devoted into the model to explore the important quantum
properties, especially the correlation function \cite{Liang}, the
entanglement \cite{nielson} and quantum phase transition
\cite{Sachdev, Peng, Quan, Cucchietti, Balian} in one dimensional
systems. The studies show that the amount of the spatial correlation
and entanglement between two spins in a spin chain (with open end)
or a spin ring (with periodical boundary condition) can be modified
by changing the magnitude of the external magnetic field and/or the
interaction strength. And these two quantities shows singularity and
obeys the scaling law in the vicinity of the quantum phase
transition point of the system. \\

The quantum phase transition, which can take place at $T=0$, is an
interesting subject and intensively studied in the last decades
\cite{Sachdev, Osterloh}. In this paper, we studied the quantum
phase transitions of the XY model through the calculation of the
correlation and the concurrence \cite{Wootters1, Wootters2} between
two nearest-neighbor spins. By changing the magnetic field or
coupling coefficient, we find that both of the two quantities have
finite jumps at some critical points. The states of the system are
carefully studied, the jumps are related to the level crossing
\cite{Zhu, Wolf, Langari} of the ground state to the excited state.
The change of the symmetry of the ground state \cite{Anderson,
Pines, Botet, Casher, Maju, Harada} is the course of the quantum
phase transition. The rest of this paper is organized as follows. In
section \ref{sec:Hamiltonian}, we introduce the Hamiltonian for the
Heisenberg XY model in an external field and describe the
computation procedures for the correlation and the concurrence of
the subsystem; Detailed results and discussions are given in section
\ref{sec:discussion}; The conclusion of our study is given in
section \ref{sec:conclusion}.

\section{The Hamiltonian and the theory}\label{sec:Hamiltonian}

The model studied here is a one dimensional spin-1/2 chain with
periodical boundary condition ({PBC}) \cite{HTWang, Marland,
Tannous} under a transverse isotropic magnetic field. The
Hamiltonian  is:
\begin{equation}\label{equ:HamiJx}
H=\sum_{i=1}^{N}(J_x\sigma_x^{(i)}\sigma_x^{(i+1)}+J_y\sigma_y^{(i)}\sigma_y^{(i+1)})
+\sum_{i=1}^{N}B_z\sigma_z^{(i)},
\end{equation}
where $N$ is the number of spins, $\sigma_x^{N+1}=\sigma_x^{1}$ and
$\sigma_y^{N+1}=\sigma_y^{1}$ from the {PBC}. $J_x$ and $J_y$ are
the coupling strength between nearest-neighbor spins along the
direction of $\vec{x}$ and $\vec{y}$, respectively, $B_z$ is the
amplitude of the magnetic field along $\vec{z}$ direction.
$\sigma_x$, $\sigma_y$ and $\sigma_z$ are the well-known Pauli
matrix:
\begin{equation}
\sigma_x=\left(\begin{array}{cc}
      0 & 1 \\
      1 & 0
    \end{array}\right), \quad
\sigma_y=\left(\begin{array}{cc}
      0 & -i \\
      i & 0
    \end{array}\right), \quad
\sigma_z=\left(\begin{array}{cc}
      1 & 0\\
      0 & -1
    \end{array}\right).
\end{equation}
The parameters $J_x$ and $J_y$ may be represented by their average value $J$ and
anisotropic parameter $\gamma$ as:
\begin{equation}\label{Jgamma2}
J_x=(1+\gamma)J, \qquad J_y=(1-\gamma)J,
\end{equation}
and the Hamiltonian may be written as:
\begin{equation}\label{equ:HamiJ}
H=2J\sum_{i=1}^{N}[(\sigma_-^{(i)}\sigma_+^{(i+1)}+\sigma_+^{(i)}\sigma_-^{(i+1)})+
\gamma(\sigma_+^{(i)}\sigma_+^{(i+1)}+\sigma_-^{(i)}\sigma_-^{(i+1)})]+
\sum_{i=1}^{N}B_z\sigma_z^{(i)},
\end{equation}
where $\sigma_{\pm}=\frac{1}{2}(\sigma_x\pm{}i\sigma_y$):
\begin{equation}
\sigma_+=\left(\begin{array}{cc}
      0 & 1 \\
      0 & 0
    \end{array}\right), \quad
\sigma_-=\left(\begin{array}{cc}
      0 & 0 \\
      1 & 0
    \end{array}\right).
\end{equation}

The spatial correlation \cite{McCoy, Schulz, Landau} between two
spins, 1 and 2, is defined by
\begin{eqnarray}\label{equ:Corre}
C_{1,2}&=&\langle\sigma^{(1)}\sigma^{(2)}\rangle
=\Tr(\rho_{12}\sigma^{(1)}\sigma^{(2)}), \\
\sigma^{(1)}\sigma^{(2)}&=&\sigma_x^{(1)}\otimes\sigma_x^{(2)}
+\sigma_y^{(1)}\otimes\sigma_y^{(2)}+
\sigma_z^{(1)}\otimes\sigma_z^{(2)}, \\ \label{equ:redrho}
\rho_{12}&=&\Tr'(\rho).
\end{eqnarray}
Where $\sigma^{(1)}\sigma^{(2)}$ can be represented explicitly in a
matrix form in the product eigenspace of $\{\sigma^2, \sigma_z\}$:
\begin{equation}
\sigma^{(1)}\sigma^{(2)}=\left(\begin{array}{cccc}
      1 & 0 & 0 & 0 \\
      0 & -1 & 2 & 0 \\
      0 & 2 & -1 & 0 \\
      0 & 0 & 0 & 1
    \end{array}\right).
\end{equation}
And $\rho_{12}$ is the reduced density matrix of the spin 1 and 2
derived from the density matrix $\rho_S$ of the whole system.
$\Tr'(\cdots)$ means to trace out all the spin degrees of freedom
except spins 1 and 2. In this study, we are interested in the ground
state of the whole system, which is the lowest energy eigenstate
$\psi_0$ of $H$, so $\rho_S=|\psi_0\rangle\langle\psi_0|$. \\

The concurrence of spins 1 and 2 is defined as \cite{Wootters1,
Wootters2}:
\begin{equation}\label{Concurrence}
Con=\max\{\lambda_1-\lambda_2-\lambda_3-\lambda_4,~0\},
\end{equation}
where $\lambda_i$ are the square roots of the eigenvalues of the
product matrix $\rho_{12}\tilde{\rho_{12}}$ in decreasing order.
Where $\rho_{12}$ is the reduced density matrix of the two spins
subsystem, and $\tilde{\rho_{12}}$ is constructed as
$(\sigma_y\otimes\sigma_y)\rho^*(\sigma_y\otimes\sigma_y)$. Equation
(\ref{Concurrence}) applies to both of the mixed state and pure
state. The reduced density matrix may be write down implicitly as:
\begin{equation}
\rho_{12}=\left(\begin{array}{cccc}
      \rho_{1,1} & \rho_{1,2} & \rho_{1,3} & \rho_{1,4} \\
      \rho_{2,1} & \rho_{2,2} & \rho_{2,3} & \rho_{2,4} \\
      \rho_{3,1} & \rho_{3,2} & \rho_{3,3} & \rho_{3,4} \\
      \rho_{4,1} & \rho_{4,2} & \rho_{4,3} & \rho_{4,4}
    \end{array}\right),
\end{equation}
then we get $\tilde{\rho_{12}}$ as the matrix below:
\begin{equation}
\tilde{\rho_{12}}=\left(\begin{array}{cccc}
      \rho_{4,4} & -\rho_{3,4} & -\rho_{2,4} & \rho_{1,4} \\
      -\rho_{3,4}^* & \rho_{3,3} & \rho_{2,3} & -\rho_{1,3} \\
      -\rho_{2,4}^* & \rho_{2,3}^* & \rho_{2,2} & -\rho_{1,2} \\
      \rho_{1,4}^* & -\rho_{1,3}^* & -\rho_{1,2}^* & \rho_{1,1}
    \end{array}\right).
\end{equation}
If the bipartite quantum state $\rho_{12}$ is pure, it can always be
written as:
\begin{eqnarray*}
\rho &=& |\psi\rangle\langle\psi|,\\
|\psi\rangle &=& a|00\rangle+b|01\rangle+c|10\rangle+d|11\rangle,
\end{eqnarray*}
then equation (\ref{Concurrence}) could be simplified to
\begin{equation}\label{Concurrence2}
Con(|\psi\rangle)=2|ad-bc|.
\end{equation}

\section{Results and discussions}\label{sec:discussion}

\begin{table}
\caption{The discontinuities in Fig. \ref{fig:SSaF4-10}}
\begin{center}
\begin{tabular}{|c|c|c|c|c|c|}
\hline  {\sl $N$} & \multicolumn{5}{c|} {\sl J (discontinuities)}
\\\hline
\hline{\sl $4$} & {\sl $0.649$} & {\sl $1.569$} & {\sl $\diagup$} & {\sl $\diagup$} & {\sl $\diagup$} \\
\hline{\sl $6$} & {\sl $0.650$} & {\sl $0.888$} & {\sl $2.426$} & {\sl $\diagup$} & {\sl $\diagup$} \\
\hline{\sl $8$} & {\sl $0.650$} & {\sl $0.767$} & {\sl $1.148$} & {\sl $3.268$} & {\sl $\diagup$} \\
\hline{\sl $10$}& {\sl $0.650$} & {\sl $0.721$} & {\sl $0.908$} & {\sl $1.415$} & {\sl $4.104$} \\
\hline
\end{tabular}
\end{center}
\label{tal:dis}
\end{table}

First, we consider the isotropic case, $\gamma=0$, in an external
field $B_z=1.30$. Figure \ref{fig:SSaF4-10} gives the correlation
functions as function of coupling strength $J$ for different sizes
of the system, ($N=4,6,8,10$). From the figures, we see that with
the increase of $J$, $C_{12}$ displays a ``ladder'' behavior.  We
also noted that for the case of $N$ spins in the system, there are
$N/2$ discontinuities, the position of the jump can be found in
table \ref{tal:dis}. \\

\begin{table}
\caption{The discontinuities of the case of $N=6$ in Fig.
\ref{fig:Cor6Jr0} and \ref{fig:Con6Jr0}}
\begin{center}
\begin{tabular}{|c|l|l|l|}
\hline  {\sl $Bz$} & \multicolumn{3}{c|} {\sl $J$ (discontinuities)}
\\ \hline
\hline{\sl 0.65=$1.30\times0.5$} & {\sl 0.325=$0.650\times0.5$} &
{\sl 0.444=$0.888\times0.5$} & {\sl 1.213=$2.426\times0.5$}  \\
\hline{\sl 1.30=$1.30\times1.0$} & {\sl 0.650=$0.650\times1.0$} &
{\sl 0.888=$0.888\times1.0$} & {\sl 2.426=$2.426\times1.0$} \\
\hline{\sl 2.60=$1.30\times2.0$} & {\sl 1.300=$0.650\times2.0$} &
{\sl 1.776=$0.888\times2.0$} & {\sl 4.852=$2.426\times2.0$} \\
\hline
\end{tabular}
\end{center}
\label{tal:disBz}
\end{table}

The effect of the magnitude of magnetic field on the phase
transition was shown in figure \ref{fig:Cor6Jr0}, in which we plot
the case of a $6$ spins system with three different magnetic fields,
$B_z=0.65$, $1.30$ and $2.6$. The corresponding transition points of
$J$ are listed in table \ref{tal:disBz}. By simple calculation and
comparison, we found that the transition values of $J$ for each
transition is approximately proportional to the magnitude of the
external field. \\

Figure \ref{fig:Cor6B} plots the variation of correlation $C_{12}$
with the applied field $B_z$ for fixed coupling coefficient $J=1.0$
and different values of anisotropic parameter $\gamma$. The system
sizes are $N=6$. When $|\gamma|<0.8$, there are $3$ jumps for a 6
spins system as in the case of isotropic case. However, with the
increase of anisotropy $\gamma$, the magnitude of the jumps becomes
smaller and the transition points moves slightly to the direction
$B_z=0$, till finally disappears in the completely anisotropic case
$\gamma=1$. Since the $\gamma=1$ corresponding to the Ising case,
where the internal number of freedom is 1, the phase transition is
different to the XY case where the internal number of freedom is 2.
Our calculation shows that the cross over from $n=2$ to $n=1$ is
continuous. It is also verified that there is no quantum phase
transition in the quantum Ising case. \\

Now we turn to the concurrence between two spins, $Con_{1,2}$. The
following results were obtained for a system of $N=6$. Figures
\ref{fig:Con6Jr0} plot the variation of concurrence $Con_{1,2}$ with
the coupling strength $J$ in the isotropic case ($\gamma=0$), the
applied magnetic field fixed to three values, $B_z = 0.65$, $B_z=
1.30$ and $B_z = 2.60$. From the figures we see the same phase
transition behavior as that of the correlation functions. Jumps are
found at exactly the same places, which can also be checked in table
\ref{tal:disBz}, as for the correlation functions. Figures
\ref{fig:Con6B} gave the variation of $Con_{1,2}$ with the magnitude
of external field $B_z$ for $8$ different anisotropy $\gamma$. When
$|\gamma|<0.5$, the jumps in concurrence are clearly observed while
for larger $|\gamma|$'s the magnitude of the jumps becomes smaller
and smaller and turn to the smooth curve in the case of $\gamma
=1$.\\

It is interesting to note that the physics behind the above phase
transitions turns out to be very simple and clear, it is the result
of level crossing of the ground state. To clarify this point, we
turn to a detailed analysis of the ground states of the system under
different sets of parameters. To be specific, we use $|1\rangle$ and
$|0\rangle$ to represent the ``spin down'' and ``spin up'' state for
each spin in the eigenspace of $\{\sigma^2, \sigma_z\}$
respectively. The state of the whole system of $N$ spins $\psi$ can
be written as a linear combination of $2^N$ states and $2^N$
normalized coefficients:
\begin{eqnarray}
\psi &=&
a_0|00\cdots00\rangle+a_1|00\cdots01\rangle+\cdots+a_n|11\cdots11\rangle,
\\ \nonumber n&=&2^N-1, \qquad \sum_{i=0}^{n}|a_i|^2 = 1.
\end{eqnarray}
where $|00\cdots00\rangle$ stands for the direct product state
$|0\rangle_1\otimes\cdots|0\rangle_N$. We diagonalized the
Hamiltonian and calculated the eigen function and the corresponding
eigen energy. We take the coupling strength $J=1$ and anisotropy
$\gamma =0$ for simplicity. Figure \ref{fig:EsBz} shows the
variation of several states energy (Someone of them will be the
ground state in some interval of $Bz$) with the magnitude of
external field $B_z$. It is clear from the figure that level
crossing occurs at the positions where abrupt changes of the
correlation function and the concurrence were observed. Concretely,
when $0<B_z\leq0.525$, the ground state (which has been normalized)
of the whole system is
\begin{equation} \label{first}
\begin{split}
\psi_0^1 =&
-0.11785|000111\rangle+0.23570|001011\rangle-0.23570|001101\rangle+
0.11785|001110\rangle \\
& -0.23570|010011\rangle+0.35355|010101\rangle-0.23570|010110\rangle
-0.23570|011001\rangle \\
&
+0.23570|011010\rangle-0.11785|011100\rangle+0.11785|100011\rangle-
0.23570|100101\rangle \\
&
+0.23570|100110\rangle+0.23570|101001\rangle-0.35355|101010\rangle+
0.23570|101100\rangle \\
& -0.11785|110001\rangle+0.23570|110010\rangle
-0.23570|110100\rangle+0.11785|111000\rangle.
\end{split}
\end{equation}
Which is basically the combination of the states with
antiferromagnetic order, that is, the states with half of the spins
``up'' and half of the spins ``down''. When
$0.526\leq{}B_z\leq1.464$, the ground state changed to
\begin{equation}\label{second}
\begin{split}
\psi_0^2=&+0.16667|000011\rangle-0.28868|001001\rangle+0.16667|001010\rangle+
0.33333|001001\rangle \\
&-0.28868|001010\rangle+0.16667|001100\rangle-0.28868|010001\rangle
+0.33333|010010\rangle \\
&-0.28868|010100\rangle+0.16667|011000\rangle+0.16667|100001\rangle-
0.28868|100010\rangle \\
&-0.28868|101000\rangle+0.16667|110000\rangle.
\end{split}
\end{equation}
This is the state that 2 spins ``up'' and 4 spins ``down''. In the
course of increasing the applied field $B_z$, the energy level
corresponding to the state (\ref{first}) increases and the energy
level corresponding to the state (\ref{second}) decreases. At the
critical field, $B_z=0.525$, the two levels meet and then crossed,
then the ground state of the spin chain is a combination of the two
level states. Beyond the critical field the ground state becomes the
state (\ref{second}). The cross of the ground state changed the
symmetry of the ground state abruptly so that a quantum phase
transition occurs. At about $B_z=1.465$, another level cross take
place and over that point, the ground state of the system becomes,
\begin{equation}\label{third}
\begin{split}
\psi_0^3=
-&0.40825|000001\rangle+0.40825|000010\rangle-0.40825|000100\rangle+
0.40825|001000\rangle \\
 -&0.40825|010000\rangle+0.40825|100000\rangle.
\end{split}
\end{equation}
This is a combination of states with 1 spins ``up'' and 5 spins
``down''. And finally, when $B_z\geq 2.001$, the ground state is
simply as
\[\psi_0^4=|000000\rangle,\]
which is the state with ferromagnetic order. We see from the above
analysis that the quantum phase transition is a result of the level
crossing of the ground state, and the level crossing is simply the
result of the competition between the coupling strength $J$, which
favors antiferromagnetic order, and the applied field, which favors
the ferromagnetic order. By increasing the applied field while
keeping the coupling strength constant, the ground state changed
from antiferromagnetic order to the ferromagnetic order in several
steps, turns a spin from ``up'' to ``down'' at each step. So we
observed that
$\psi_0^1\rightarrow\psi_0^2\rightarrow\psi_0^3\rightarrow\psi_0^4$.
This also explains the number of jumps in the correlation function
and concurrence, which must be half of the whole number of spins.

\section{Conclusion}\label{sec:conclusion}

In this paper, we studied the spatial correlation $C_{1,2}$ and the
degree of entanglement $Con_{1,2}$ between two nearest-neighbor
spins along a one-dimension spin ring in Heisenberg XY model. We
found that in the isotropic case with fixed external field $B_z$,
quantum phase transitions of both $C_{1,2}$ and $Con_{1,2}$ were
observed with the increment of coupling strength $J$, the phase
transitions will smoothed out as the anisotropy $\gamma$ increased
and disappears at the extremum value of $\gamma=1$. The transition
is a direct result of the level crossing of the ground state of the
whole system, in which the symmetry of the ground state changed
abruptly at the crossing point.

\newpage
\section*{Figure Captions}

\begin{figure}[htbp]
  \centering
\includegraphics[scale=0.7]{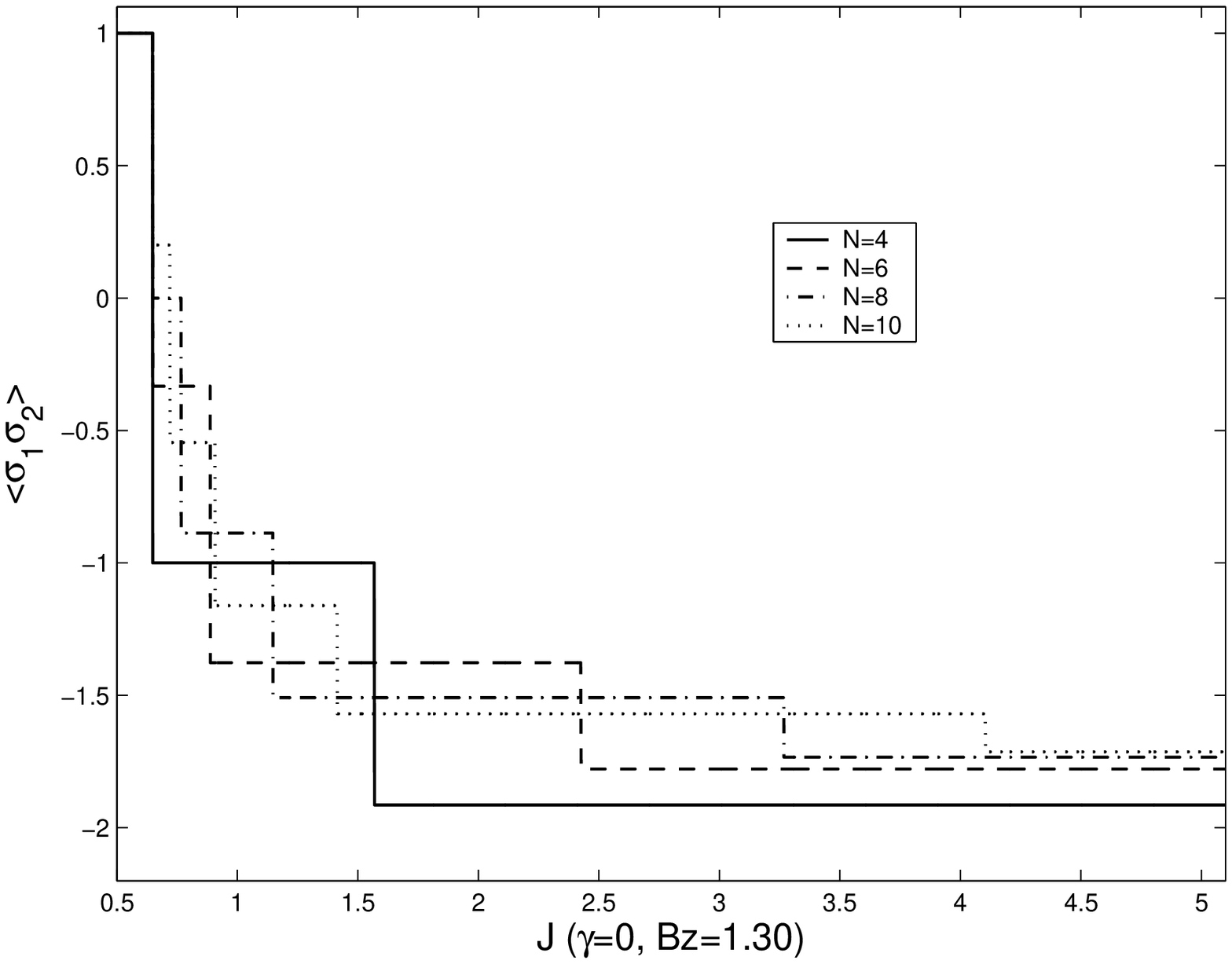}\\
  \caption{} \label{fig:SSaF4-10}
\end{figure}

FIG. \ref{fig:SSaF4-10}: The  nearest-neighbor correlation $C_{1,2}$
in different size of the spin ring as function of coupling strength
$J$, parameters are
$\gamma=0$ and $Bz=1.30$.\\

\begin{figure}[htbp]
  \centering
\includegraphics[scale=0.7]{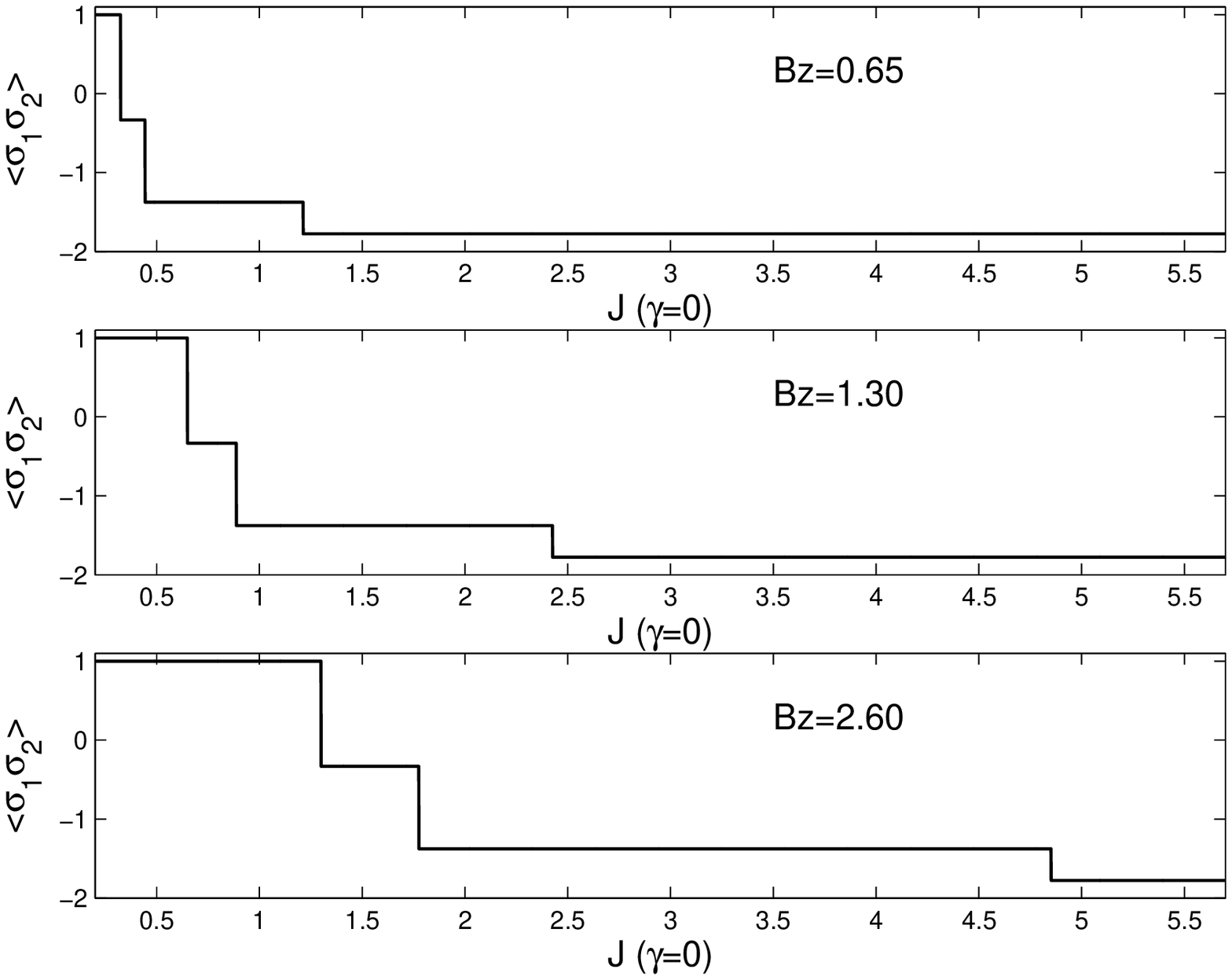}\\
  \caption{} \label{fig:Cor6Jr0}
\end{figure}

FIG. \ref{fig:Cor6Jr0}: The  nearest-neighbor correlation $C_{1,2}$
under three  different cases of external field: $B_z=0.65$,
$B_z=1.30$ and $B_z=2.60$, the parameter $\gamma=0$. \\

\begin{figure}[htbp]
\centering \subfigure[$J=1.0,\gamma=0$]{ \label{fig:Cor6B:0}
\includegraphics[width=2.5in]{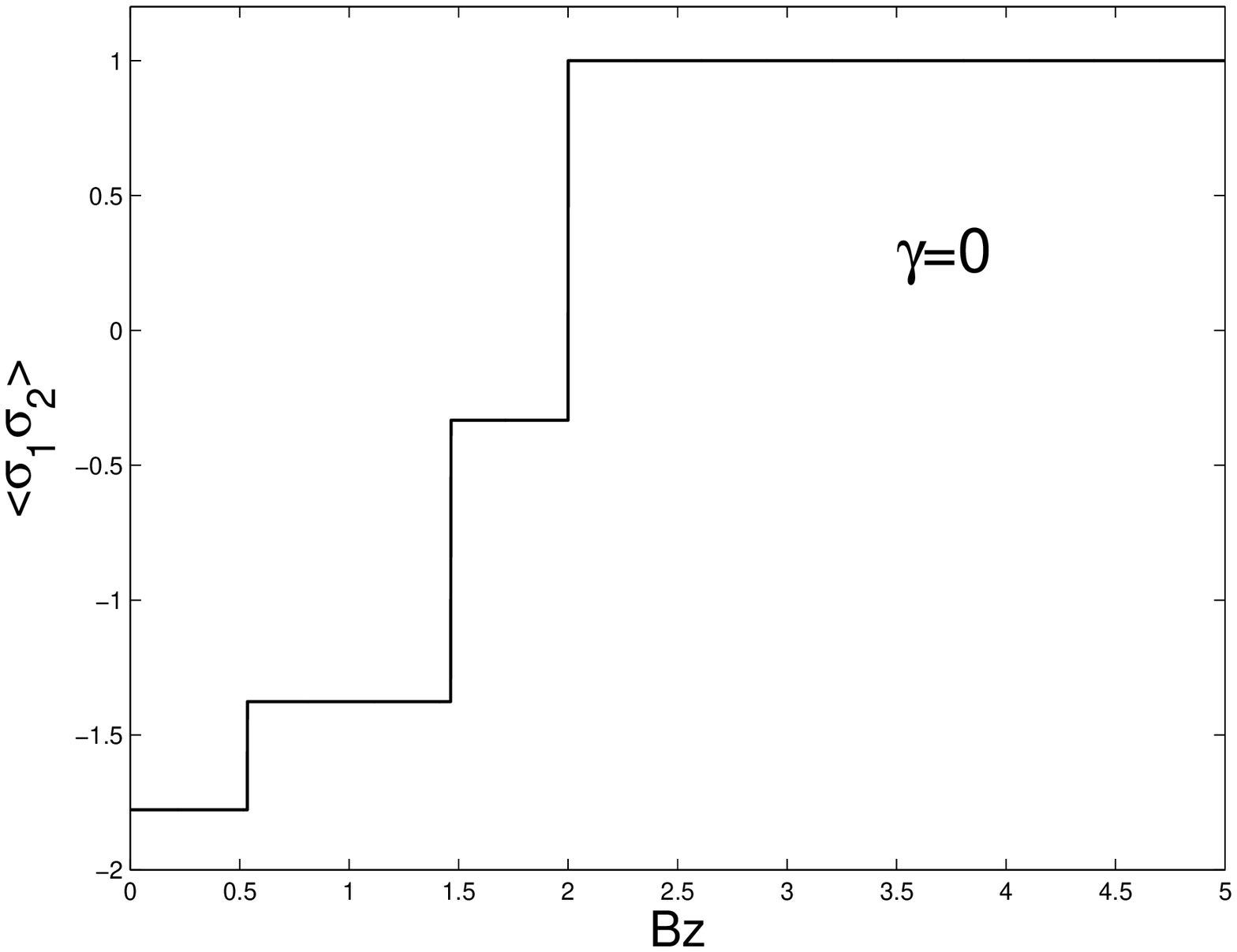}}
\subfigure[$J=1.0,|\gamma|=0.1$]{ \label{fig:Cor6B:0.1}
\includegraphics[width=2.5in]{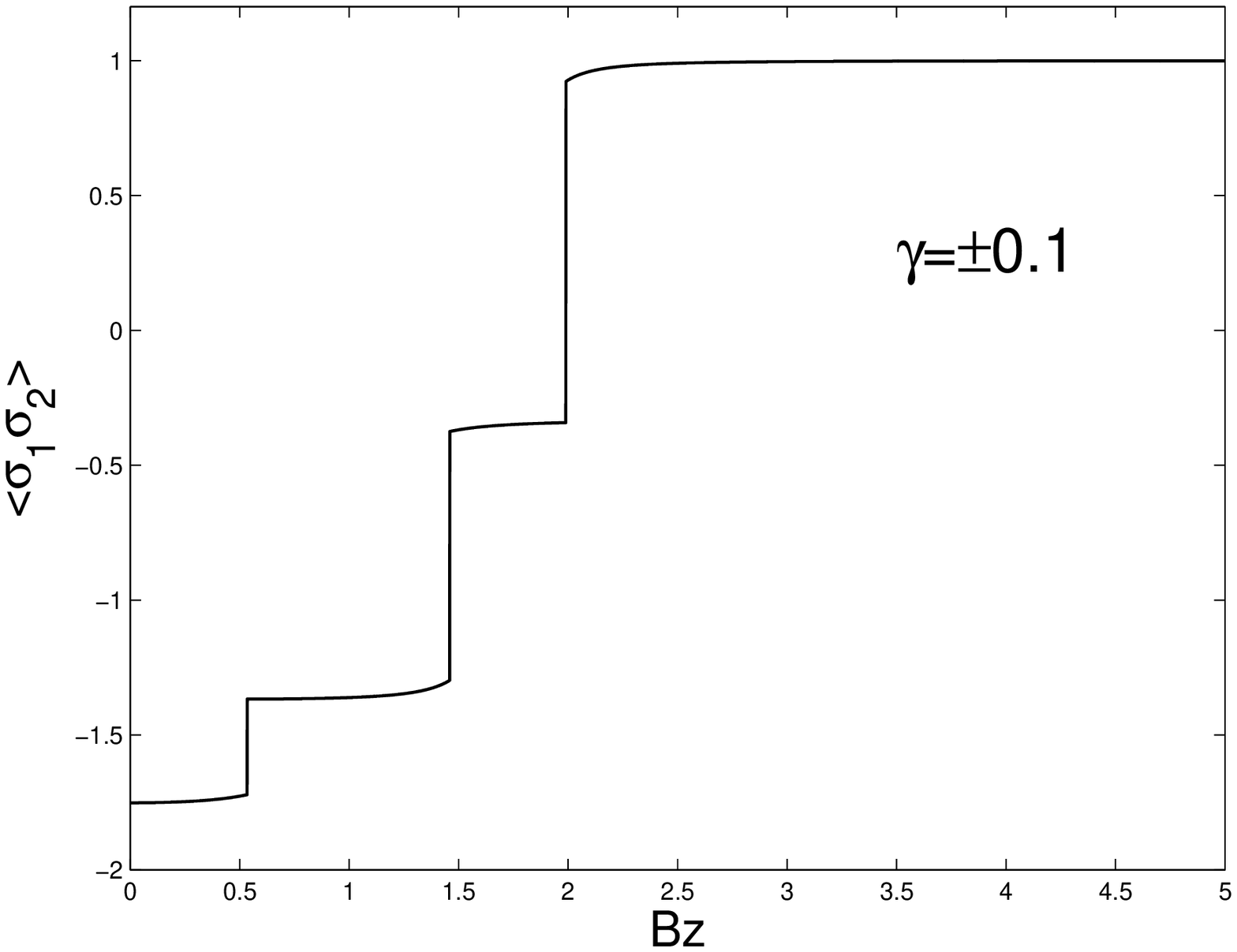}}
\\
\subfigure[$J=1.0,|\gamma|=0.3$]{ \label{fig:Cor6B:0.3}
\includegraphics[width=2.5in]{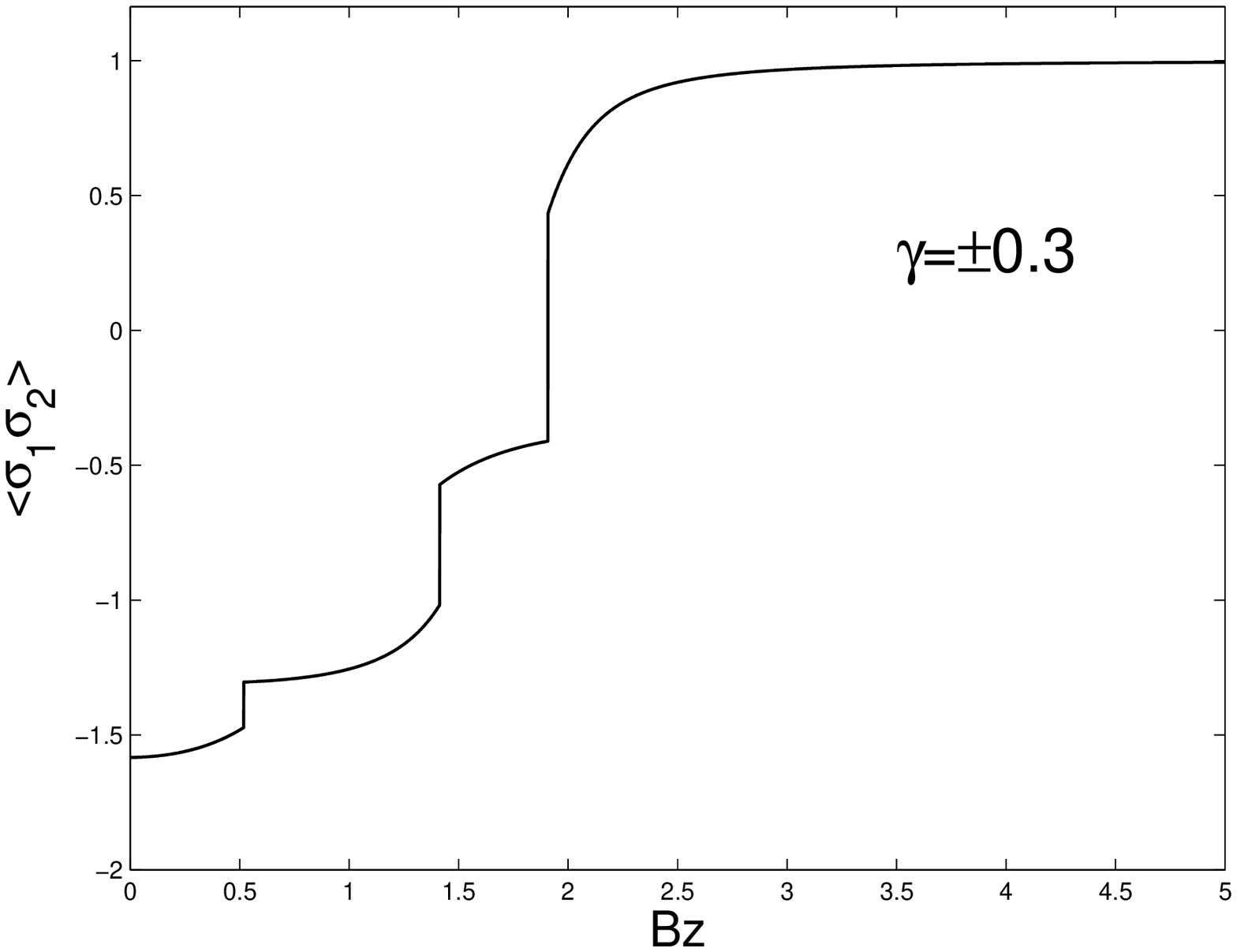}}
\subfigure[$J=1.0,|\gamma|=0.4$]{ \label{fig:Cor6B:0.4}
\includegraphics[width=2.5in]{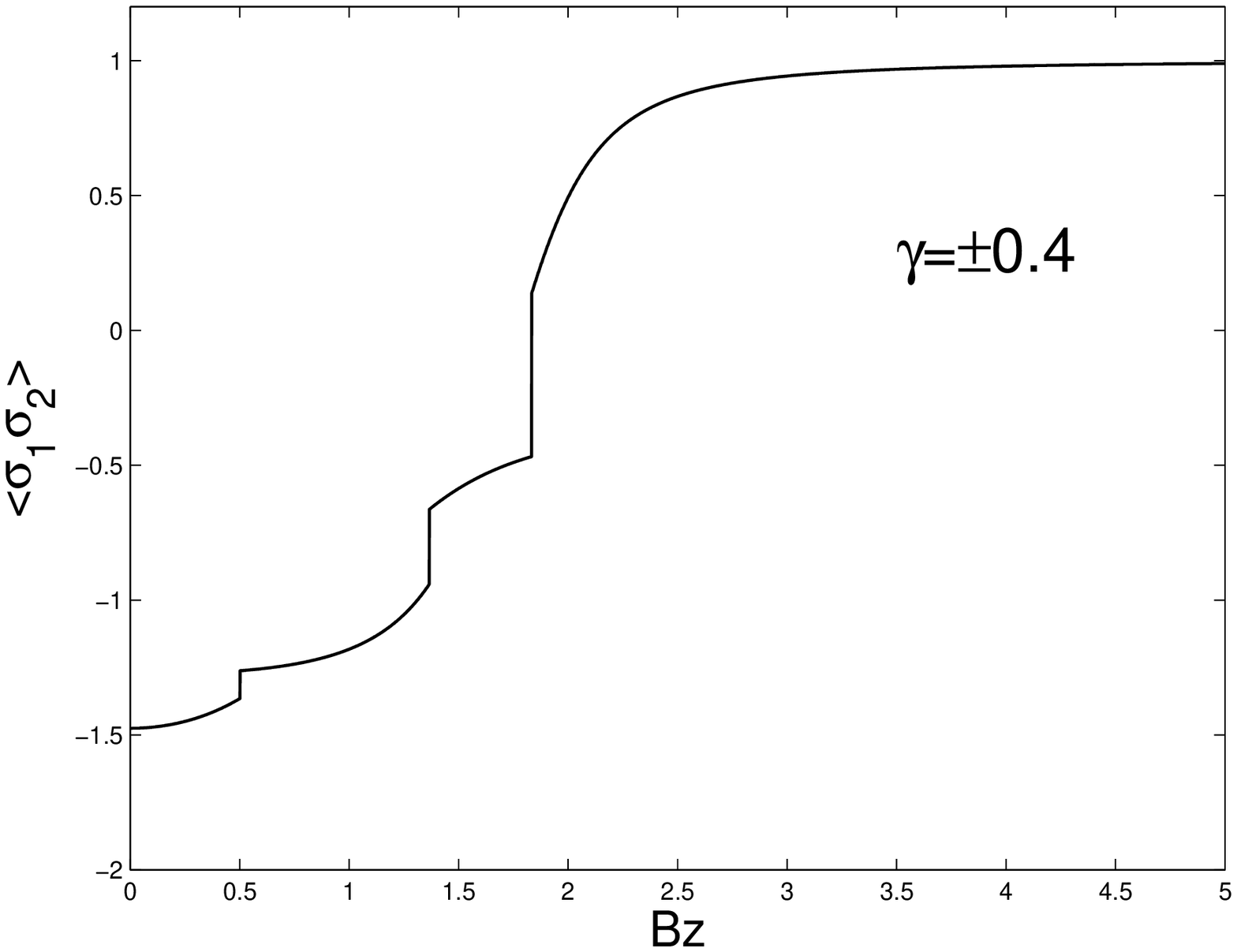}}
\\
\subfigure[$J=1.0,|\gamma|=0.5$]{ \label{fig:Cor6B:0.5}
\includegraphics[width=2.5in]{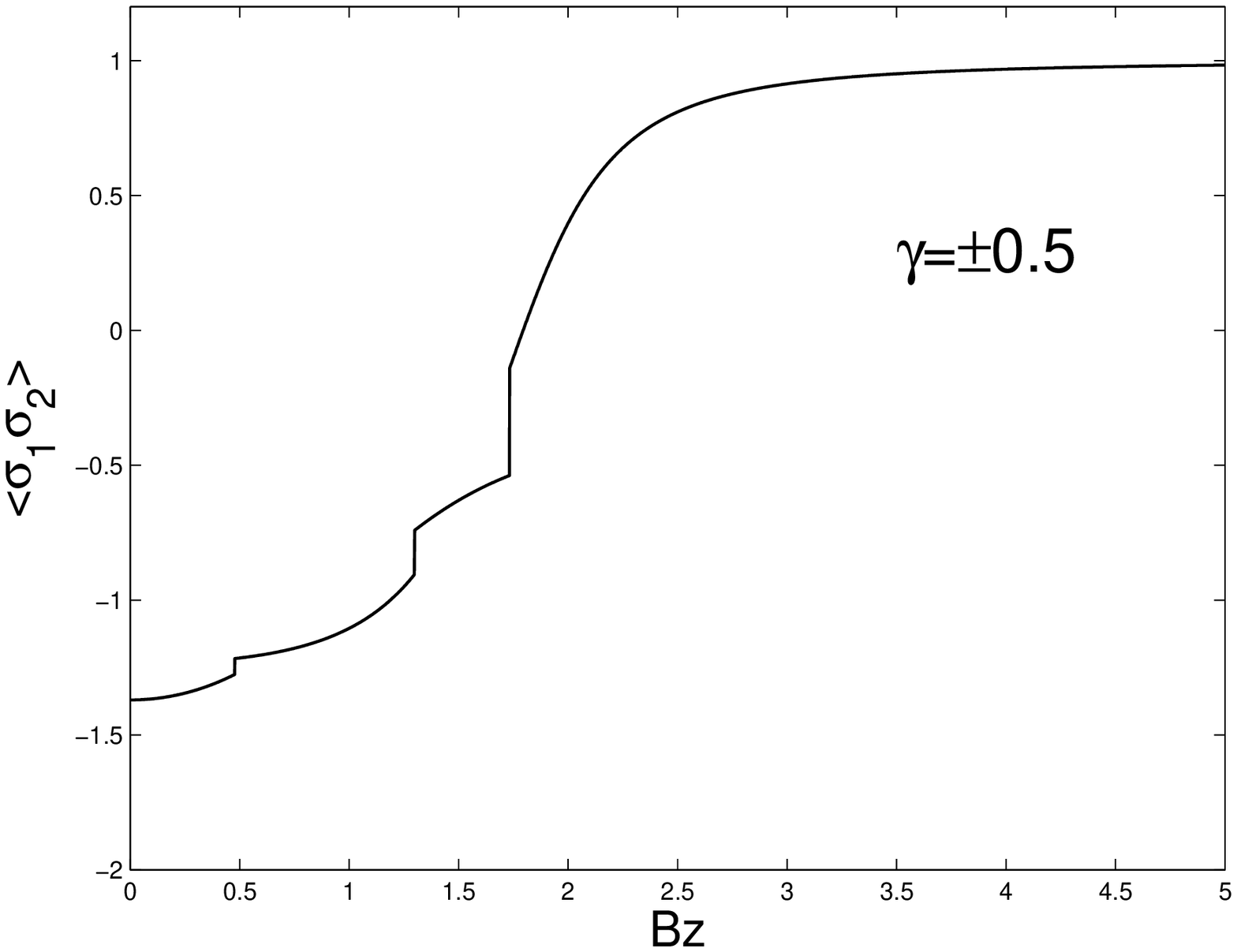}}
\subfigure[$J=1.0,|\gamma|=0.6$]{ \label{fig:Cor6B:0.6}
\includegraphics[width=2.5in]{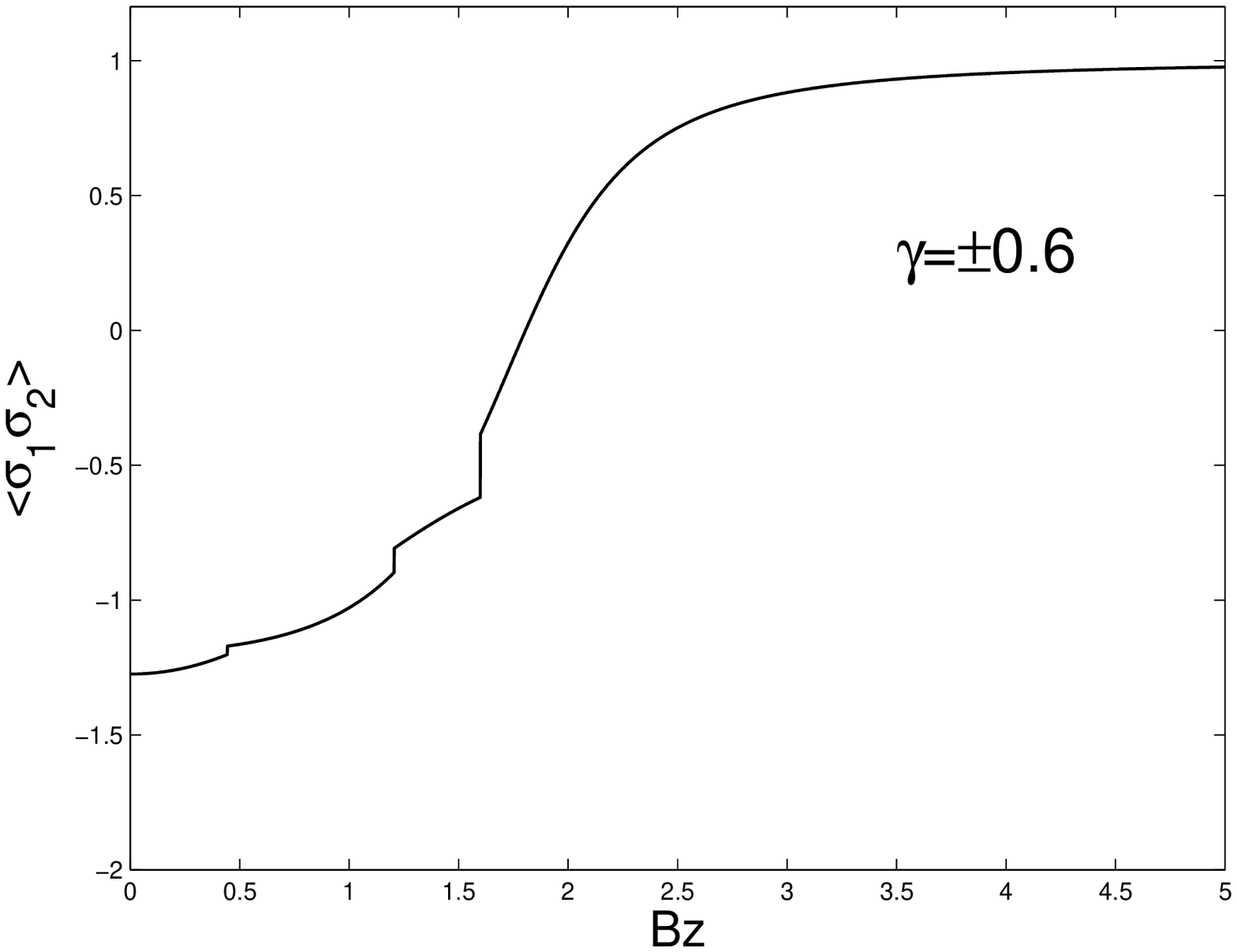}}
\\
\subfigure[$J=1.0,|\gamma|=0.8$]{ \label{fig:Cor6B:0.8}
\includegraphics[width=2.5in]{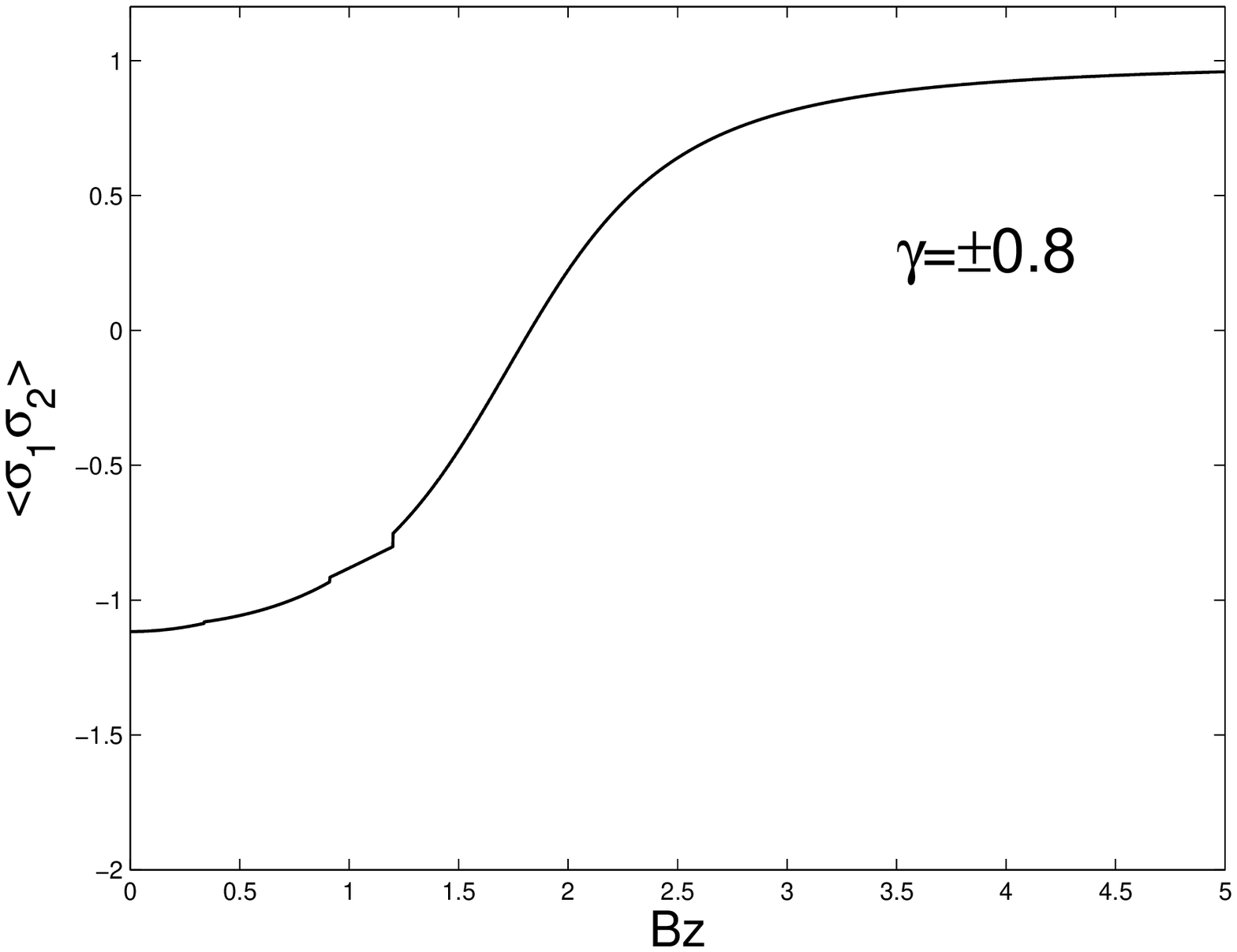}}
\subfigure[$J=1.0,|\gamma|=1.0$]{ \label{fig:Cor6B:1}
\includegraphics[width=2.5in]{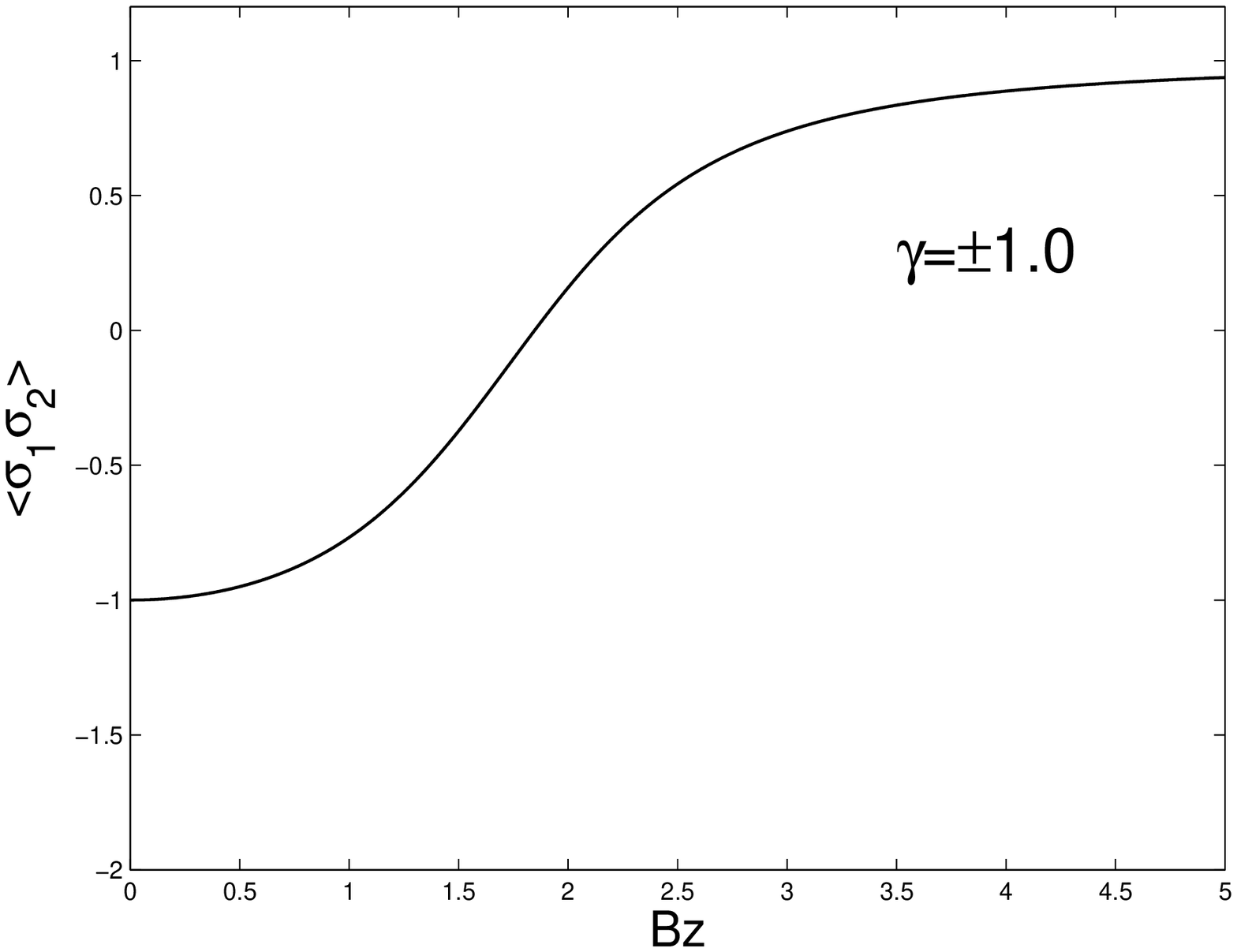}}
\caption{} \label{fig:Cor6B}
\end{figure}

FIG. \ref{fig:Cor6B}: The dependence of nearest-neighbor correlation
$C_{1,2}$ on the applied field $B_z$ with different anisotropic
parameters $\gamma$. All the results
are calculated with  $J=1.0$.\\

\begin{figure}[htbp]
  \centering
\includegraphics[scale=0.7]{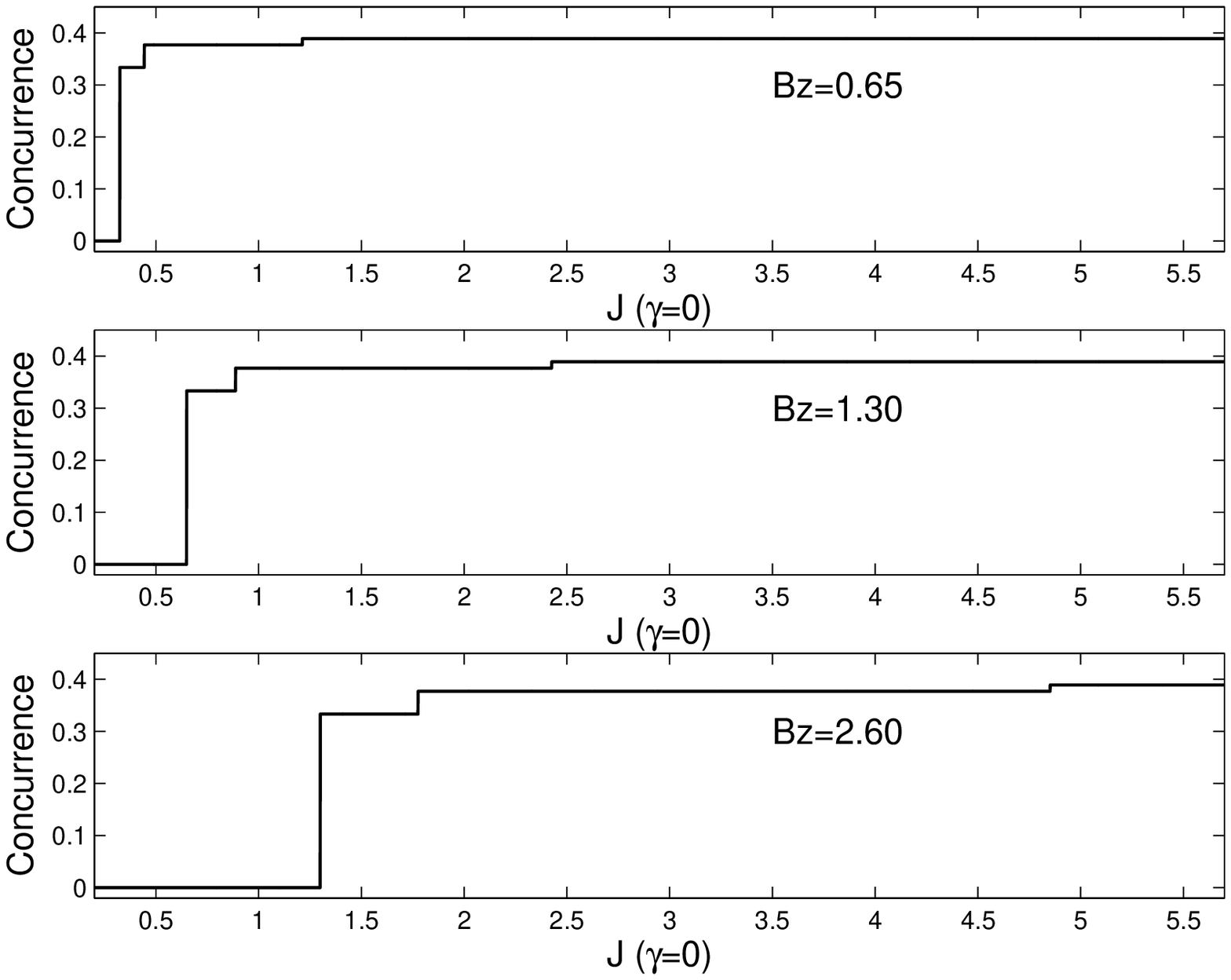}\\
  \caption{} \label{fig:Con6Jr0}
\end{figure}

FIG. \ref{fig:Con6Jr0}: The comparison of Concurrence for two spins
in  the  isotropic case, $\gamma=0$,  in three  different applied
field: $B_z=0.65$, $B_z=1.30$ and
$B_z=2.60$.\\

\begin{figure}[htbp]
\centering \subfigure[$J=1.0, \gamma=0$]{ \label{fig:Con6B:0}
\includegraphics[width=2.5in]{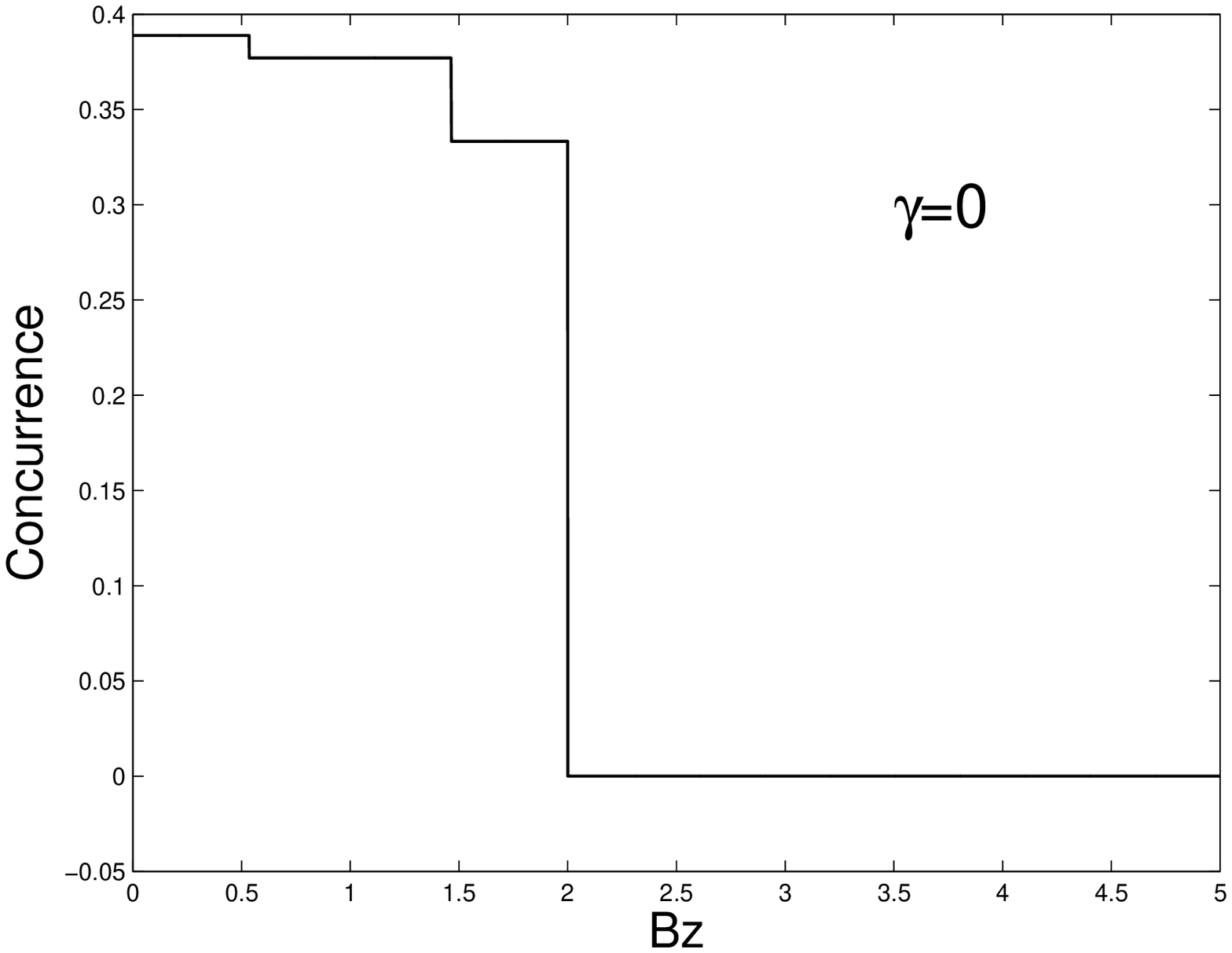}}
\subfigure[$J=1.0, |\gamma|=0.1$]{ \label{fig:Con6B:0.1}
\includegraphics[width=2.5in]{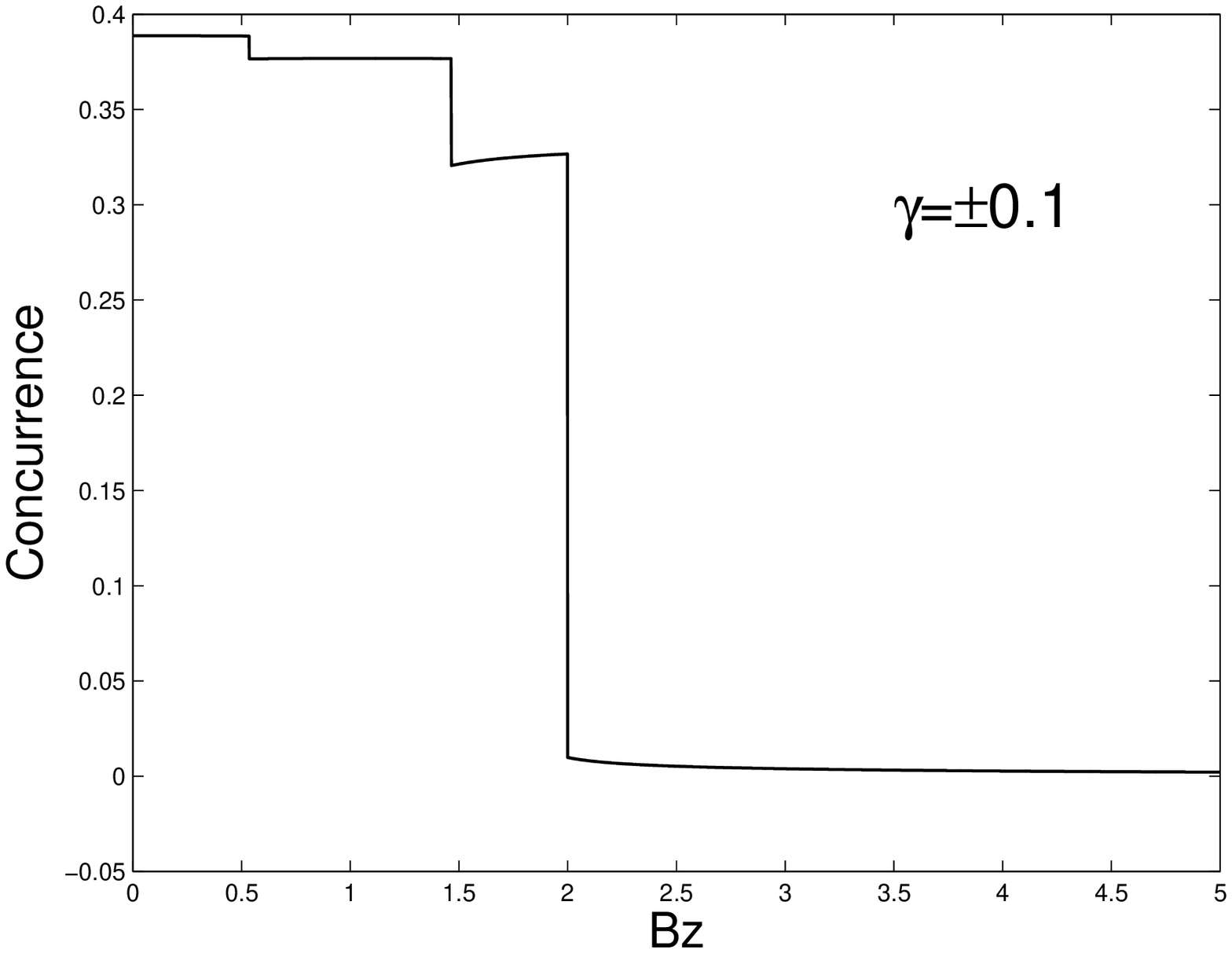}}
\\
\subfigure[$J=1.0, |\gamma|=0.3$]{ \label{fig:Con6B:0.3}
\includegraphics[width=2.5in]{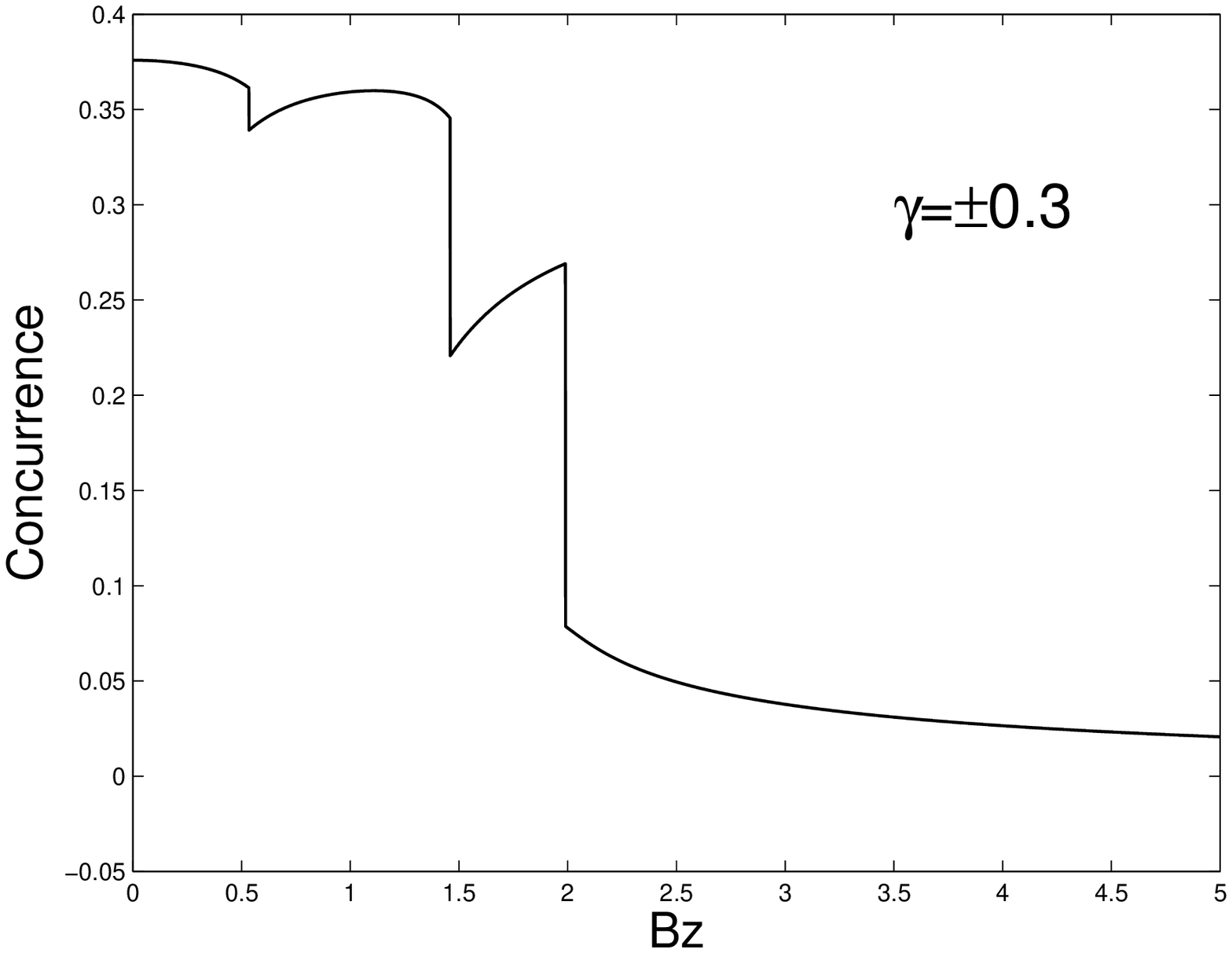}}
\subfigure[$J=1.0, |\gamma|=0.4$]{ \label{fig:Con6B:0.4}
\includegraphics[width=2.5in]{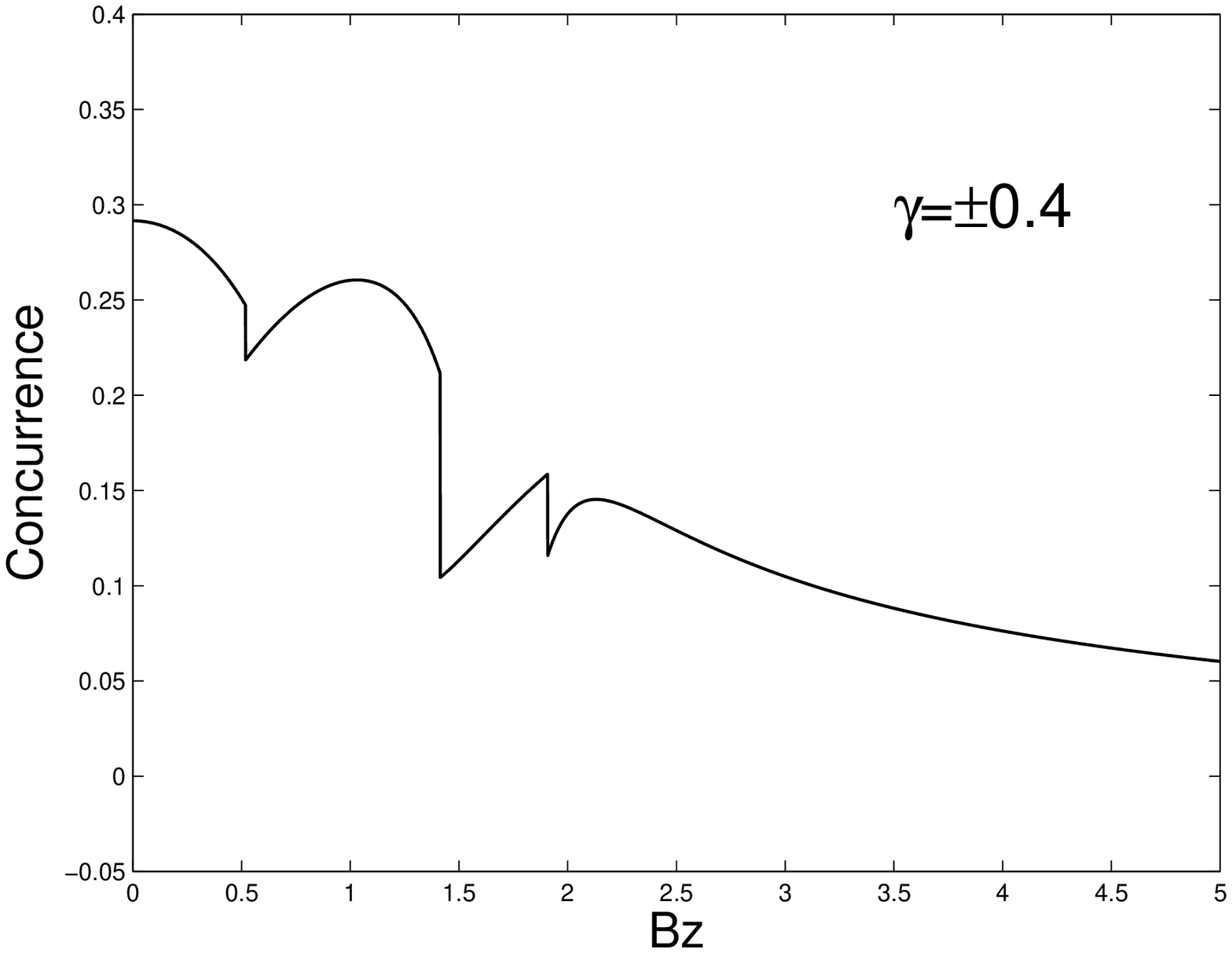}}
\\
\subfigure[$J=1.0, |\gamma|=0.5$]{ \label{fig:Conr6B:0.5}
\includegraphics[width=2.5in]{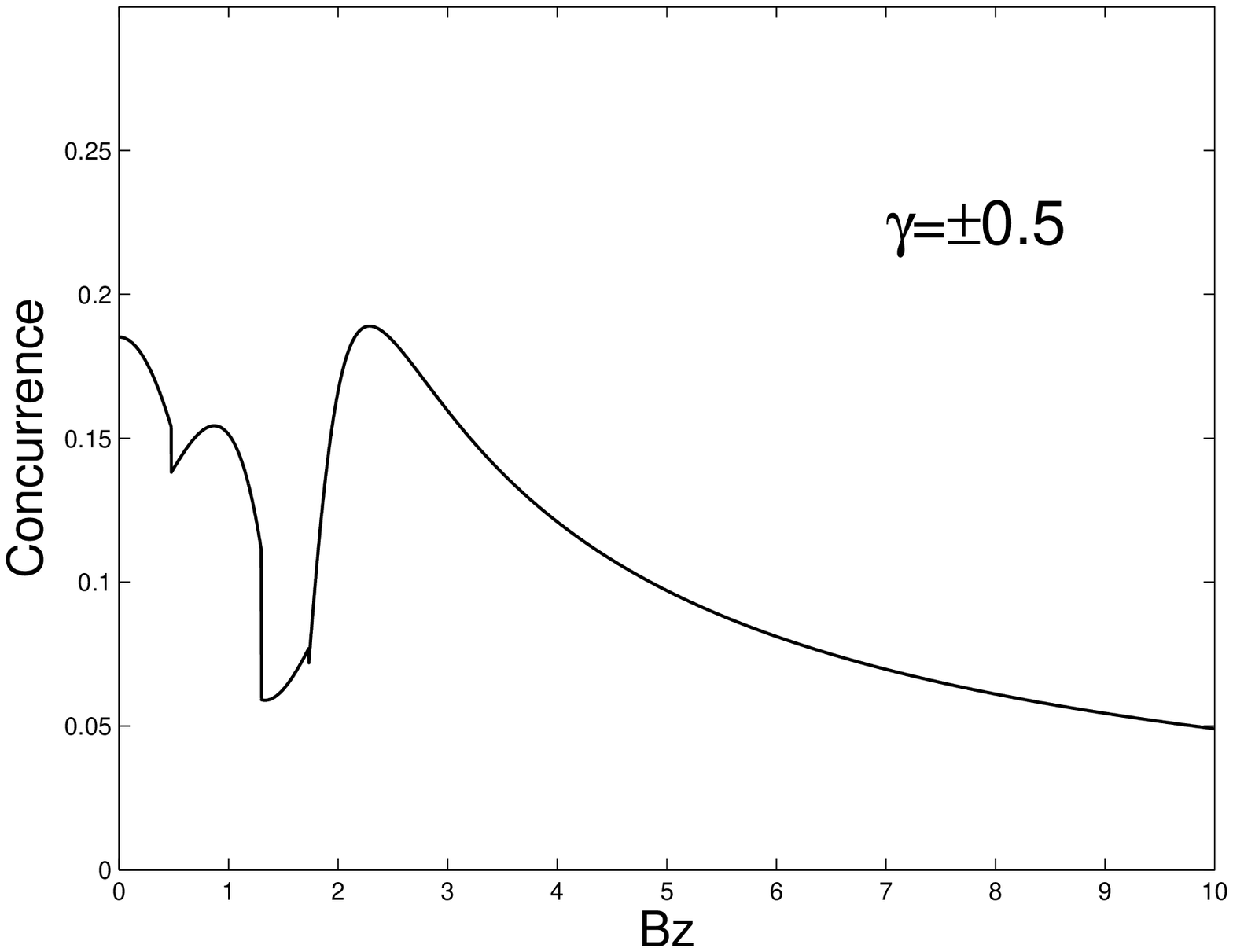}}
\subfigure[$J=1.0, |\gamma|=0.8$]{ \label{fig:Con6B:0.8}
\includegraphics[width=2.5in]{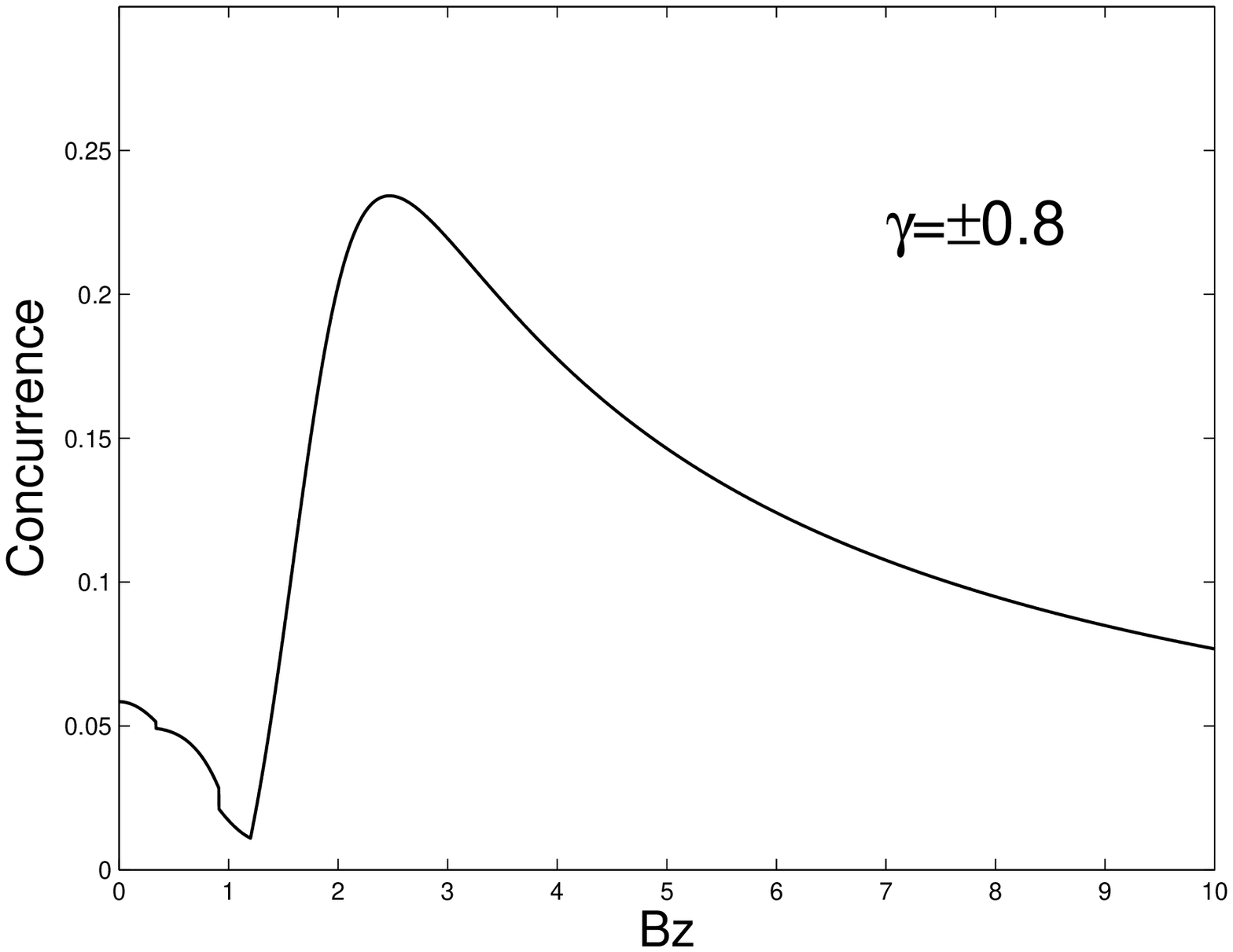}}
\\
\subfigure[$J=1.0, |\gamma|=0.9$]{ \label{fig:Con6B:0.9}
\includegraphics[width=2.5in]{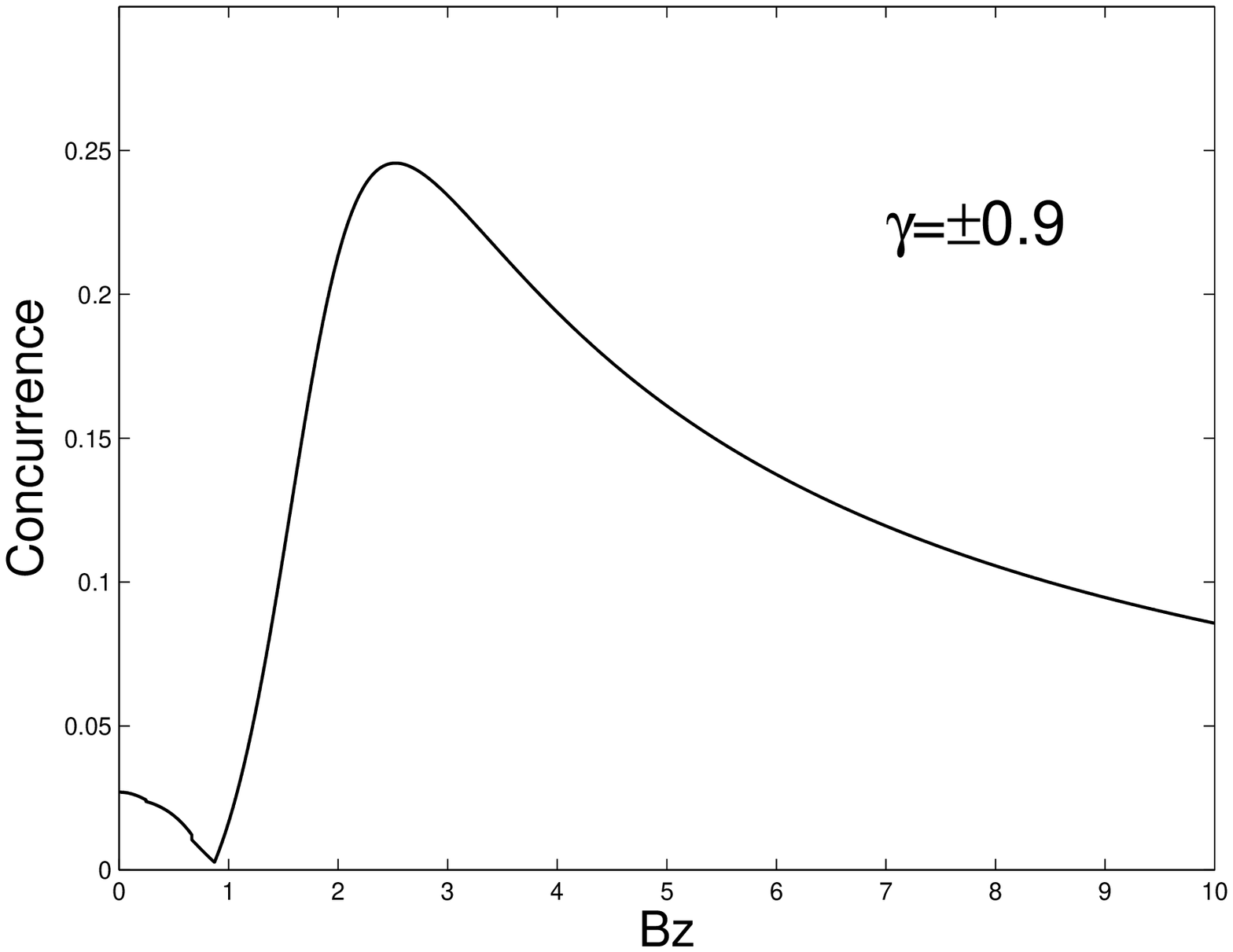}}
\subfigure[$J=1.0, |\gamma|=1.0$]{ \label{fig:Con6B:1}
\includegraphics[width=2.5in]{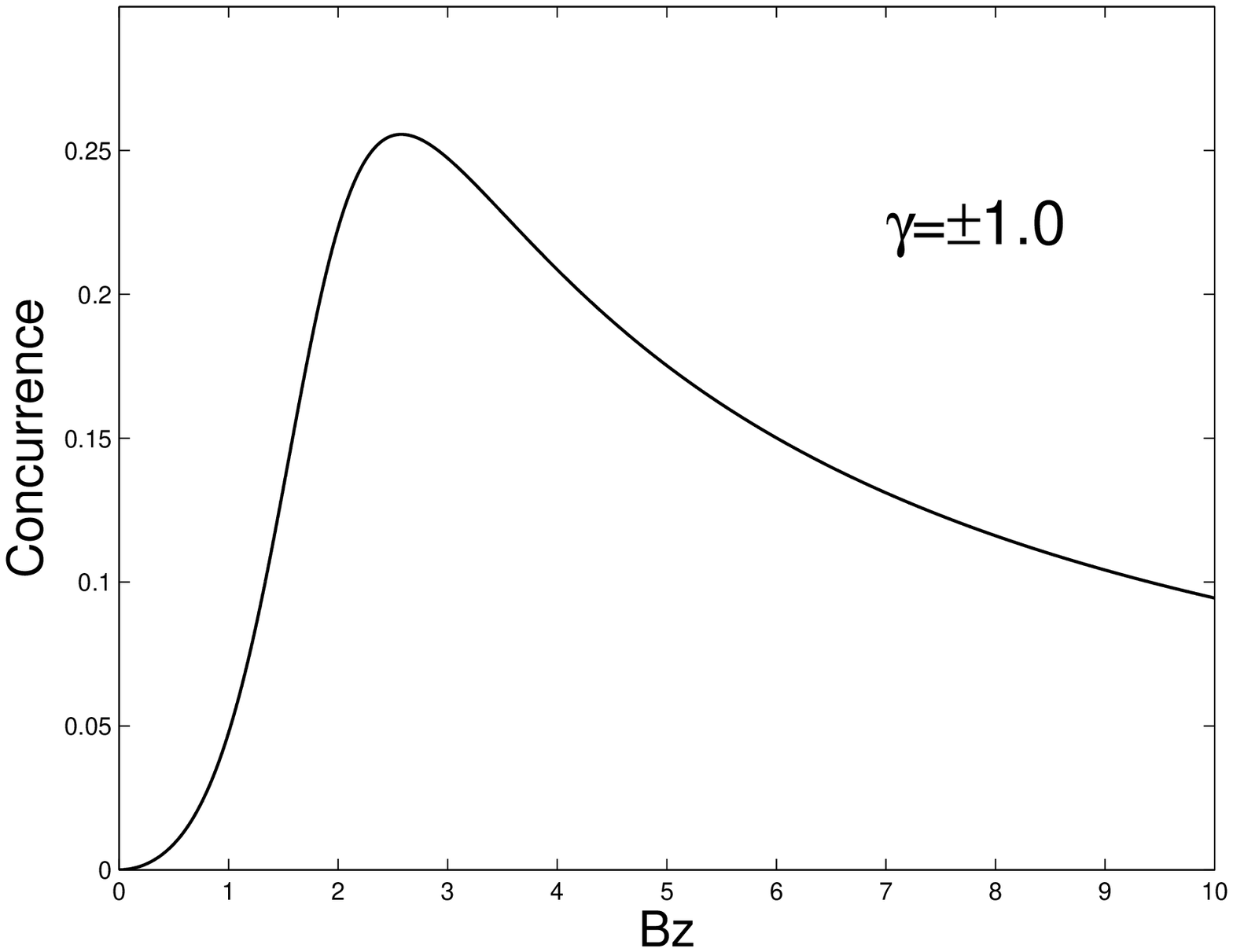}}
\caption{} \label{fig:Con6B}
\end{figure}

FIG. \ref{fig:Con6B}: The dependence of Concurrence Concurrence for
two spins on the applied field $B_z$ with different anisotropic
parameters $\gamma$. All the results are calculated with
 $J=1.0$.\\

\begin{figure}[htbp]
\includegraphics[width=10cm]{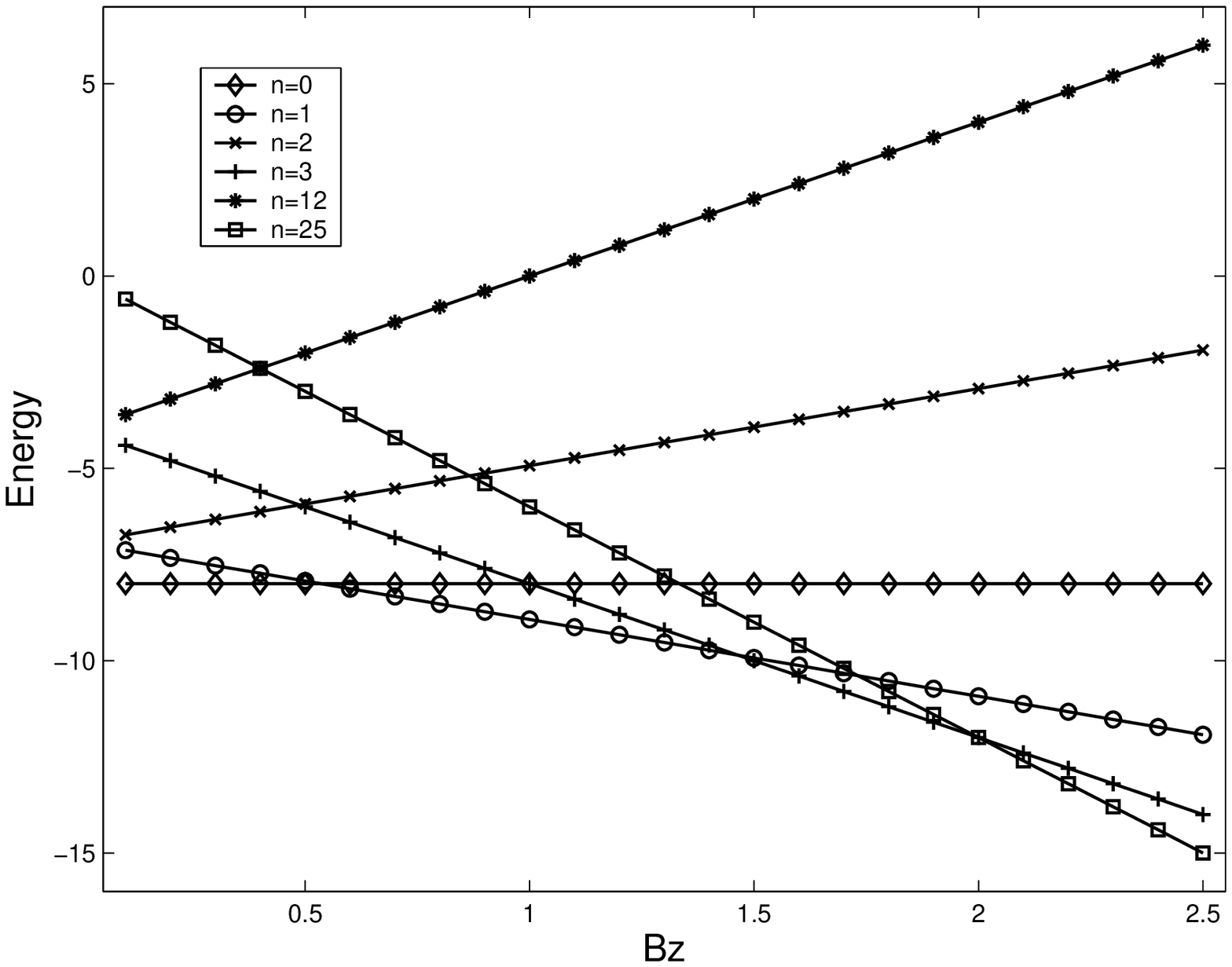}
\caption{}\label{fig:EsBz}
\end{figure}

FIG. \ref{fig:EsBz}: The energy levels as function of $B_z$ for
given coupling strength, the system size is $N=6$, some of the
levels which do not take part in the level crossing
were omitted for clarity. The other parameters are: $\gamma=0.0$,  $J=1.0$.\\

\newpage

\newpage

\newpage

\newpage

\newpage

\newpage
\end{document}